\documentstyle[11pt]{article}

\newcommand{\EQ}{\begin{equation}}
\newcommand{\EN}{\end{equation}}
\begin{document}
\topmargin 0pt
\oddsidemargin=-0.4truecm
\evensidemargin=-0.4truecm
\renewcommand{\thefootnote}{\fnsymbol{footnote}}
%\newpage
\setcounter{page}{1}
\begin{titlepage}
\begin{flushright}
\Large
INFN FE-08-93 \\
June 1993
\end{flushright}
\vspace{0.7cm}
\begin{center}
{\Large MATTER INDUCED NEUTRINO DECAY: \\
\vspace{0.2cm}
NEW CANDIDATE FOR THE SOLUTION TO THE SOLAR \\
\vspace{0.2cm}
NEUTRINO PROBLEM }\footnote{Talk presented at $5^{th}$ Int. Workshop on
 "Neutrino Telescopes", Venice, March 2-4, 1993. To appear \\
on the Proceedings.}
\vspace{1.2cm}

{\Large Zurab G. Berezhiani}\footnote{E-mail: 39967::berezhiani,
 berezhiani@ferrara.infn.it,
zurab@hep.physik.uni-muenchen.de}\\
\vspace{0.2cm}
{\em Istituto Nazionale di Fisica Nucleare, sezione di Ferrara,I-44100
Ferrara,
Italy\\
Institute of Physics, Georgian Academy of Sciences, Tbilisi
380077, Georgia}\\
\vspace{0.4cm}
{\Large Anna Rossi}\footnote{E-mail: 39967::rossi, rossi@ferrara.infn.it } \\
\vspace{0.2cm}
{\em Dipartimento di Fisica, Universita' di Ferrara, I-44100 Ferrara,
Italy\\ Istituto Nazionale di Fisica Nucleare, sezione di Ferrara, I-44100
Ferrara,
Italy}\\

\end{center}
\begin{abstract}
\Large
Effects of the dense matter can induce the neutrino decay even if neutrinos
are stable in vacuum. This is due to coherent interactions with matter
which lead to energy level-splitting between the neutrino and
antineutrino states and thereby provide
available phase space for the majoron emission.
We show that the matter induced decay
can be a plausible candidate for the explanation of solar neutrino deficit,
provided that there exist new flavour-changing weak range interactions of
neutrino with matter constituents, and constant of $\tau$-neutrino
coupling to the majoron is sufficiently large.
This mechanism naturally implies the hierarchy
between the ratios $Z=\frac{measured\,signal}{SSM\,expectation}$ for Ga-Ge,
$\nu_ee$ scattering
and Cl-Ar experiments - $Z_{GA}>Z_{K}>Z_{Cl}$. The important feature of
the matter induced decay is the prediction of solar antineutrino flux with
substantially degraded energy spectrum compared to solar neutrinos.
This scenario can be unambiguously tested by future solar neutrino
detectors BOREXINO/BOREX and SNO.
\end{abstract}
\end{titlepage}
\renewcommand{\thefootnote}{\arabic{footnote}}
\setcounter{footnote}{0}
\newpage
\Large
{\bf 1. Solar Neutrino Problem (SNP).} The discrepancy between the
neutrino flux predicted by the Standard Solar Model (SSM) \cite{B,BP}
and the observed flux, is the main open issue in neutrino physics.
The ratio $Z$ of the observed signal to that is expected from the SSM
is different for different solar neutrino experiments. Namely,  the
Homestake Cl-Ar experiment gives \cite{Davis}
\EQ
Z_{Cl}=0.28 \pm 0.04
\EN
%where the error is from $R_{obs}$(Ar) only.
The results of Kamiokande II and Kamiokande III are \cite{Hirata}:
\EQ
Z_{KII}=0.46\pm 0.05\pm 0.06, ~~~~~ Z_{KIII}=0.56\pm 0.07\pm 0.06
\EN
The combined result of both implies $Z_{K}=0.49\pm 0.05\pm 0.06$.
% \cite{Lang}.%for recoil electrons of energy larger than 7.5 MeV.
Finally, the results of the Ga-Ge radiochemical experiments GALLEX
\cite{Gallex} and SAGE \cite{Sage} are
\EQ
Z^{GALLEX}_{Ga}=0.63\pm 0.14\pm 0.06, ~~~~~~~~
Z^{SAGE}_{Ga}= 0.44\pm 0.15\pm 0.11
\EN
with the combined result $Z_{Ga}= 0.54\pm 0.11$ (all the above
data are given with $1\sigma$ error due to experiment and do not include
"$3\sigma$" theoretical uncertainties of the SSM \cite{BP}).
Focusing on the central values and
%If one takes seriously the whole set of these results,
neglecting experimental uncertainties, the following hierarchy of
the data is obeyed:
\EQ
Z_{Ga}> Z_{K}> Z_{Cl}
\EN
This points out that the SNP cannot be explained by a non-standard
temperature of the solar core, since this would imply $Z_{Cl}>Z_{K}$
\cite{Lang}.

The non-standard particle physics solutions require generally the
violation of lepton number or lepton flavour, that leads to new neutrino
properties and new physical phenomena.
Namely, neutrinos produced in the sun can be converted into neutrinos of
different flavour and/or helicity while they propagate to the earth.
The popular explanations of the solar neutrino deficit through the neutrino
oscillations in vacuum\cite{Pont,BPW1} or in matter \cite{MSW,MS} and
spin-flavour transitions \cite{VVO,ALM} are still waiting for
being confirmed (or excluded) by new data from future
facilities.\footnote{The averaged vacuum oscillations
$\nu_e\rightarrow\nu_{\mu},\nu_{\tau}$ predict, in contradiction with the
data, the equal signal for two radiochemical experiments:
$Z_{Ga}=Z_{Cl}>0.33$, and also $Z_{K}=Z_{Cl}+0.15(1-Z_{Cl})>Z_{Ga}$.
However, the vacuum oscillations in "Just So" regime \cite{BPW1} still
can be regarded as a plausible candidate for the SNP solution \cite{BPW2}.}

{\bf 2. The SNP solution through neutrino decay in vacuum.}
The idea that the deficit of solar neutrinos can be due to their decay
during the flight from sun to earth was suggested long time ago \cite{BCY}.
Since the fast radiative decay is excluded both from
particle physics and astrophysical arguments, one has to consider fast
invisible decay modes of the neutrino, e.g. with the emission of majoron,
the Goldstone boson related to spontaneous violation of the global
lepton number symmetry $U(1)_{B-L}$. From the viewpoint of the majoron
model building the possibility of neutrino decay during the flight time
$t\simeq 500$ s implies the following two conditions:

 (i) {\em Sufficiently strong $\nu$-majoron couplings} ($h>10^{-4}$).
This in turn requires very low scale of the lepton number
violation ($\eta_{BL}<10$ keV).
The most familiar candidate for such a low $\eta_{BL}$,
the triplet majoron model \cite{GR}, has been ruled out by LEP data
on $Z$-boson invisible width, whereas the "seesaw" type singlet majoron
\cite{CMP} generally implies $\eta_{BL}>100$ GeV and therefore is
extremely weakly coupled to neutrinos. However, a variety of new singlet
majoron models can be considered \cite{BSV,Burgess} in which the
scale $\eta_{BL}$ can be sufficiently low as to provide
coupling constants in the needed range.\footnote{Another implication of
these models can be observable neutrinoless $2\beta$ decay with majoron
emission.}

(ii) {\em Existence of majoron off-diagonal couplings.}
As it was emphasized in \cite{Joe}, the neutrino decay scenario cannot be
realized in simple majoron models, in which global $U(1)_{B-L}$ symmetry
acts on all lepton families in the same way. In order to achieve the
existence of majoron tree-level off-diagonal couplings between neutrino mass
eigenstates one has to complicate the theory.
Namely, different lepton flavours should
be distinguished by different charges of the global $U(1)$ symmetry
(in which case the "virgin" idea of the $U(1)_{B-L}$
symmetry is actually lost), or different lepton number symmetries
($U(1)_e \otimes U(1)_{\mu} \otimes U(1)_{\tau}$)
should be invoked \cite{Jose}.

The simplest decay scenario with negligible neutrino mixing, i.e.
the case $\nu_e\rightarrow\nu_x+\chi$ with decay length adjusted to the
sun-earth distance \cite{BCY,Betal}, is completely excluded by the
$\bar{\nu}_e$ pulse observation from SN1987A. However, this
does not rule out the scenario with large neutrino
mixing \cite{BV,Haber}. Its implications were
investigated to a full extent in \cite{BFMR,AIP}. It was shown that this
scenario can reconcile the Davis and Kamiokande data, leading to
$Z_{Cl}<Z_K$. However, this scenario implies $Z_{Ga}<Z_{Cl}$ due to the
energy dependence of the decay probability in vacuum
which suppresses more the low energy neutrinos. Thus, this mechanism is
disfavoured by the GALLEX data even if not excluded yet.

{\bf 3. Matter Induced Decay (MID).}
As it was shown in \cite{BV}, the effects of dense matter can induce
the neutrino decay with majoron emission even in the case of simplest
majoron model, when neutrinos are stable in vacuum. The point is that
the coherent interactions with medium lead to the energy splitting
between the $\nu$ and $\bar{\nu}$ states providing available phase space
for the emission of the majoron. Subsequently the implications of neutrino
decay in matter were studied in a number of papers \cite{BS,Kim,Choi,GKL}.

Let us remind  the main features of the MID with majoron emission.
For the simplicity we consider the case of two neutrino
flavours $\nu_e$, $\nu_x$ ($x=\mu,\tau$),
which are defined as left-handed Weyl spinors: $\nu_{e}=\nu_{eL}$,
$\nu_{x}=\nu_{xL}$ (the antineutrino states have opposite chirality:
$\tilde{\nu}_{e}=\tilde{\nu}_{eR}=C\bar{\nu}^{T}_{eL}$,
 $\tilde{\nu}_{x}=\tilde{\nu}_{xR}=C\bar{\nu}^{T}_{xL}$,
where C is the matrix of charge conjugation).
In the majoron picture the neutrino masses and mixing arise from the Yukawa
couplings to some complex scalar field
$\sigma$ with non-zero vacuum expectation value
$\langle\sigma\rangle=\eta_{BL}/\sqrt{2}$, which spontaneously violates
the lepton number:
\EQ
\sigma=\frac{1}{\sqrt{2}}(\eta_{BL}+\rho)e^{i\chi}
\EN
where $\rho$ is a Higgs scalar with a mass $\sim \eta_{BL}$ and $\chi$ is
a massless majoron. In the flavour basis the  Lagrangian of neutrino
interaction with the majoron has the form
\EQ
L=(\bar{\nu}_{e},\bar{\nu}_{x})\left(
\begin{array}{cc}
h_{ee} & h_{ex}\\
h_{xe} & h_{xx}
\end{array}\right)\frac{i}{2}\chi
\left(
\begin{array}{c}
\tilde{\nu}_{e}\\ \tilde{\nu}_{x}
\end{array}\right)+h.c.
\EN
and the neutrino mass matrix reads as
\EQ
\hat{M}=\left(\begin{array}{cc}
m_{ee} & m_{ex}\\
m_{xe} & m_{xx}
\end{array}\right)=\eta_{BL}\left(\begin{array}{cc}
h_{ee} & h_{ex}\\
h_{xe} & h_{xx}
\end{array}\right)
\EN
Obviously, the matrix (6) becomes diagonal together with (7).
Thus the majoron couplings with neutrino eigenstates
$\nu_{1}=c\nu_{e}+s\nu_{x},~~\nu_{2}=-s\nu_{e}+c\nu_{x}$, with masses
$m_1$ and $m_2$ respectively, are the following:
\EQ
L  = (\bar{\nu}_{1},\bar{\nu}_{2})\left(
\begin{array}{cc}
h_{1} & \\
 & h_{2}
\end{array}\right)\frac{i}{2}\chi
\left(
\begin{array}{c}
\tilde{\nu}_{1}\\ \tilde{\nu}_{2}
\end{array}\right)+h.c.
\EN
where
\begin{displaymath}
c = \mbox{cos}\theta,~~ s=\mbox{sin}\theta,
{}~~~~\mbox{tg}2\theta=2h_{ex}/(h_{xx}-h_{ee})
\end{displaymath}
\EQ
h_{1}\,=\,c^{2}h_{ee}+s^{2}h_{xx}+2csh_{ex},~~~~~
h_{2}\,=\,c^{2}h_{xx}+s^{2}h_{ee}-2csh_{ex}
\EN

As far as there are no off-diagonal tree-level $\nu$-majoron couplings,
 the heavier neutrino mass eigenstate cannot decay into the lighter one
with majoron emission. Thereby, in the simplest majoron models the
neutrinos are stable in vacuum \cite{Joe}.

However, the presence of matter will induce the neutrino decay even in the
simplest majoron model. Indeed, the neutrino propagation in matter is
described by the following Schr\"odinger equation:
\EQ
i\frac{d}{dt}\left(
\begin{array}{c}
\nu_{e}\\ \nu_{x}
\end{array}\right)
=\hat{H}_{\nu}\left(
\begin{array}{c}
\nu_{e}\\ \nu_{x}
\end{array}\right)
\EN
where the Hamiltonian reads:
\EQ
\hat{H}_{\nu}=\left(\begin{array}{cc}
V_e+(c^{2}m^{2}_1+s^{2}m^{2}_2)/2E & -cs(m^{2}_2-m^{2}_1)/2E \\
-cs(m^{2}_2-m^{2}_1)/2E & V_x+(c^{2}m^{2}_2+s^{2}m^{2}_1)/2E
\end{array}\right).
\EN
where E is the neutrino energy
and $V_{e}$ and $V_{x}$ are the matter induced potentials
of the current eigenstates $\nu_{e}$ and $\nu_{x}$ respectively:
\EQ
V_{e,x}=\sqrt{2}G_{F}\frac{\rho}{m_{N}}v_{e,x}\,;~~~~~ v_e=
Y_{e}-\frac{Y_{n}}{2}=1 -\frac{3}{2}Y_n\,,  ~~~~~~ v_x=-\frac{Y_n}{2}.
\EN
Here $G_{F}$ is the Fermi constant, $\rho$ is the matter density, $m_{N}$
is the nucleon mass, and $Y_{e,n}$ are the
number of electrons and neutrons per nucleon ($Y_e=1-Y_n$ for the
electrically neutral matter).
The evolution of antineutrino states is described by the matrix
$\hat{H}_{\tilde{\nu}}$ of form analogous to (11) but with potentials
of opposite sign: $V_{\tilde{e},\tilde{x}}=-V_{e,x}$. Therefore, the
$\nu-\tilde{\nu}$ level-splitting appears that
provides non-zero phase space for certain transitions between
neutrino and antineutrino
matter eigenstates with majoron emission. Clearly, the vacuum (mass)
eigenstates do not coincide
with the matter eigenstates. Moreover, in this case of light
enough neutrinos ($m^{2}\ll VE$) the latter are essentially the
flavour eigenstates $\nu_e, \nu_x$ (and $\tilde{\nu_e}, \tilde{\nu}_x$ for
antineutrinos), so that the $\nu-\tilde{\nu}$ transition matrix
is given by (6) and $\hat{H}_{\tilde{\nu}}=-\hat{H}_{\nu}$. Therefore, the
majoron transitions can be flavour diagonal as well as flavour
changing (but necessarily with helicity-flipping). The corresponding
decay widthes were calculated in \cite{BV}:
\EQ
\Gamma_{e\tilde{e}}= \frac{h^{2}_{ee}}{16\pi}2V_e, ~~~~~~~
\Gamma_{e\tilde{x}}= \frac{h^{2}_{ex}}{16\pi}(V_e+V_x), ~~~~~
\Gamma_{x\tilde{x}}= \frac{h^{2}_{xx}}{16\pi}2V_x
\EN
where $\Gamma_{e\tilde{e}}\equiv \Gamma (\nu_e\rightarrow\tilde{\nu}_e +
 \chi)$ etc. Negative width means that the corresponding $\nu$ and
$\tilde{\nu}$ states must be interchanged (e.g. since $V_x< 0$,
$\tilde{\nu}_x$ decays into $\nu_x$).
These decay widthes do not depend on neutrino energy, because the increase
 of the phase space for a fast moving neutrino ($\propto EV$) is cancelled
 by time dilatation effect ($\propto 1/E$). The energy distribution of the
secondary states does not depend on matter potentials \cite{BS}:
\EQ
W(E,E')= \frac{1}{\Gamma(E)} \, \frac{d\Gamma(E,E)}{dE'}= 2 \,
\frac{E- E'}{E^{2}}
\EN
where E' is the energy of the secondary antineutrino.
Therefore, the secondary $\tilde{\nu}$ is strongly degraded - in average
2/3 of the initial neutrino energy E is taken away by the majoron.

The probability that the neutrino will undergo  MID passing a
medium with varying density does not depend on its energy.
The flux of $\nu_e$ survived the MID at the distance R from the origin, is
\cite{BS}
\EQ
\Phi_e(R) = \Phi^{0}_e
\exp\left(-\frac{1}{8d_0}\sum_{x}h^{2}_{ex}d^{eff}_{ex}\right),
\EN
where $d_{0}=\sqrt{2}\pi G^{-1}_{F}m_{N}\simeq 1.6\cdot 10^9$ g/cm$^2$
is the refraction width and
\EQ
d^{eff}_{ex}=\int_{0}^{R}\rho(r) \left[v_{e}(r)+v_{x}(r)\right]dr,
{}~~~~~~~x=e, \mu, \tau.
\EN
 According to eqs. (12) we have:
\EQ
d^{eff}_{ee}=2d^{eff}-3d^{eff}_n,~~~~~~~
d^{eff}_{ex}=d^{eff}-2d^{eff}_n~~~(x=\mu,\tau)
\EN
where
\EQ
d^{eff}=\int_{0}^{R}\rho(r)dr,~~~~~d^{eff}_n=\int_{0}^{R}Y_n(r)\rho(r)dr
\EN
are the matter effective widthes traversed by neutrinos.

The properties of the MID are drastically different from the properties
of decay in vacuum. First, the MID of neutrino occurs into the state of
opposite helicity,
i.e. antineutrino state, whereas in vacuum both the helicity conserving
and helicity flipping modes have comparable decay width.
Second, in matter both flavour-changing
$\nu_e\rightarrow\tilde{\nu}_{\mu,\tau}+\chi$
and flavour conserving $\nu_e\rightarrow\tilde{\nu}_e+\chi$ decays are
possible. Third, the MID exhibites unusual dependence
of neutrino lifetime on its energy in laboratory frame, which does not
depend or even decreases with energy,
quite opposite to the case of decay in vacuum when the slow particles
decay faster.
Finally, the neutrino lifetime versus the decay in matter is
effectively determined by the matter width passed by neutrinos.

Recalling that for the solar medium $Y_n\leq 0.33$, the solar neutrinos can
decay in both channels, $\nu_e\rightarrow\tilde{\nu}_e+\chi$ and
$\nu_e\rightarrow\tilde{\nu}_x+\chi$ ($x=\mu,\tau$).
Let us calculate the mean matter widthes  $<d^{eff}>$
for the solar neutrinos coming
from different sources ($^{8}B, ^{7}Be, pp$), averaged with their
production distribution in the sun and different directions of flight:
\EQ
<d^{eff}>=\frac{1}{4\pi}\int_{V}\int_{\vec{n}} d\vec{n}
d\vec{r} Q(\vec{r})\tilde{d}^{eff}(\vec{r}, \vec{n})
%}{\int_{V}\int_{\vec{n}} d\vec{n} dV Q(V)}
\EN
where $d\vec{r}$ is the elementary volume, $d\vec{n}$ is the elementary
solid angle, $Q(\vec{r})$ are the distributions of production rates for
the neutrinos from different sources and $\tilde{d}^{eff}$ is the matter
width passed by neutrino created in the volume
$d\vec{r}$ and flying in the direction $d\vec{n}$. The above expression is
easily simplified due to spherical symmetry of the sun -
 $Q(\vec{r})= \frac{1}{4\pi r^2}Q(r)$:
%were the distributions $Q(r)$ are tabulated in \cite{B}
\EQ
<d^{eff}>= \int_{0}^{R}\int_{0}^{1}dr d(\mbox{cos}\theta')Q(r)
\tilde{d}^{eff}(r\mbox{sin$\theta'$})
\EN
were $\theta'$ is the angle between $\vec{r}$ and $\vec{n}$, and
\EQ
\tilde{d}^{eff}(r\mbox{sin$\theta'$})=\int_{0}^{R}dl\rho(z), ~~~~~
z=\sqrt{r^2\mbox{sin}^{2}\theta'+l^2}
\EN
The expression for the other width $<d^{eff}_n>$ is completely the same
with $\rho\rightarrow\rho Y_n$ in eq. (21).
Taking the distribution functions $Q(r)$ for the neutrinos from different
sources ($^{8}B, ^{7}Be, pp$) as they are tabulated in \cite{B}, we have
(in g/cm$^2$):
\begin{displaymath}
<d^{eff}> = 1.38\cdot 10^{12},~~~~~
<d^{eff}_n> =   0.29\cdot 10^{12}~~~~~~~
(^{8}B: \, E=0-15\, MeV)
\end{displaymath}
\begin{displaymath}
<d^{eff}> = 1.26\cdot 10^{12},~~~~~
<d^{eff}_n> =   0.25\cdot 10^{12}~~~~~~~
(^{7}Be: \, E=0.861\,MeV)
\end{displaymath}
\EQ
<d^{eff}> = 1.08\cdot 10^{12},~~~~~
<d^{eff}_n> =   0.20\cdot 10^{12}~~~~~~~
(pp: \, E=0-0.42\,MeV)
\EN
These differences are due to the fact that high energy neutrinos are mostly
produced in the deeper and more dense solar core and thereby have to pass
larger matter width before leaving the sun.

Therefore, if the solar neutrino deficit can be related with MID, then
the relation $Z_{Ga}>Z_{Cl}$ is expected naturally. This is due to the
fact that the solar $pp$ neutrinos (which do not contribute to the Chlorine
experiment, but are responsible for about 55\% of the signal in Gallium
experiments) pass about 20-30\% less effective width compared to the
Boron neutrinos and thereby have less chance to undergo MID.

However, in the case of Hamiltonian (11), due to
%as it was shown in \cite{BV,BMR},
only standard interactions (i.e. neutrino scattering off the particles
with Z and W boson exchange), the effect
of MID with majoron emission cannot provide a solution to the
SNP due to the strong existing bounds on the $\nu$ -majoron coupling
costants:
\begin{displaymath}
h_{ee} < 3\cdot10^{-4}~~~ \cite{Moe},~~~~~~~ \\
\end{displaymath}
\EQ
\sum_{x}h_{ex}^{2}\leq 4.5\cdot 10^{-5}\,~~~~~
\sum_{x}h_{\mu x}^{2}\leq 5.4\cdot 10^{-4}~~~~
(x=e,\mu,\tau,...)~~~\cite{BKP}
\EN
 Moreover, these constraints allow
$\tilde{\nu_{e}}$ signal originated by the solar neutrino
decay to be at most at the borderline of detectability
even for the future large volume detectors like Super-Kamiokande or BOREX
\cite{BMR} - only few percents of
solar neutrinos can undergo MID and, moreover, the energy spectrum of
secondary antineutrinos is strongly degraded. In the case when neutrino masses
are not negligible, the decay probabilities become even less \cite{BMR}.
%The fact that $\nu_{\tau}$-majoron coupling $h_{\tau\tau}$ can be large,

However, the $\nu_{\tau}$-majoron coupling constant $h_{\tau\tau}$ is not
really restricted by any laboratory constraints. It can be as large as
$O(10^{-1})$, providing very fast decay
$\tilde{\nu}_{\tau}\rightarrow\nu_{\tau}+\chi$ in the solar medium.
This cannot solve SNP by the simple reason that
the solar neutrinos are not $\tilde{\nu}_{\tau}$.
However, the presence of neutrino non-standard weak range interactions
with matter constituents can drastically change the situation.
 In the next section we shall show how to take advantage of this fact.

{\bf 3. MID due to non-standard neutrino interactions.} Indeed, the MID
scenario can be more appealing if possible non-standard neutrino
interactions with the matter particles are included.
Generally, such interactions emerge inevitably in the context of singlet
majoron models \cite{BSV} that potentially can provide the $\nu$-majoron
coupling constants in a strong regime relevant for MID. These models
utilize new charged (or coloured+charged) scalars with masses within 100
GeV range. Their exchange, after proper Fierz trasformation,
effectively provides the new channels of the neutrino neutral vector current
scattering with quarks and charged leptons,  which
effectively contribute to the neutrino potentials in unpolarized medium
(see also majoronless models of refs.\cite{Valle2,Roulet,BPW3}).
These non-standard interactions generally can be flavour-conserving
as well as flavour changing.\footnote{It was shown in refs.
\cite{Roulet,BPW3} that such non-standard interactions of neutrinos,
for a certain region of corresponding coupling constants,
can effectively induce resonant neutrino conversion in solar medium
even in the absence of neutrino mass terms and thereby solve the SNP.}

 Let us
analyse the possible impact of such interactions in the presence of enough
strong $\nu_{\tau}$ - majoron coupling. Taking into account that the
$\mu$-neutrino cannot be relevant due to the strong bound
(23) on $\nu_{\mu}$-majoron coupling constants, we omit for the
simplicity the $\nu_{\mu}$ state and assume that $\nu_x=\nu_{\tau}$. Let us
consider, as an example,  only neutrino elastic scattering off d-quarks
(e.g. in the context of "coloured" Zee model in ref. \cite{BSV}) due to
the following NC interactions:
\EQ
L_{eff}=-\sum_{\alpha,\beta}\sqrt{2}G^{d}_{\alpha\beta} \,
(\bar{\nu}_{\alpha L}\gamma_{\mu}\nu_{\beta L}) \,
(\bar{d}\gamma_{\mu}d+\xi_{\alpha\beta}\bar{d}\gamma_{\mu}\gamma^5d),
{}~~~~(\alpha,\beta=e,\tau)
\EN
Bearing in mind that only vector currents are relevant for the coherent
neutrino scattering off unpolarized medium, we define 3 new parameters
which represent the ratios of the new amplitudes to the standard one
\EQ
\varepsilon_{e,\tau}=
A^{VNC}(\nu_{e,\tau}d\longrightarrow\nu_{e,\tau}d)/A^{W} =
G^{d}_{ee,\tau\tau}/G_{F}
\EN
\EQ
\varepsilon_{e\tau}= A^{VNC}(\nu_{e}d\longrightarrow\nu_{\tau}d)/A^{W} =
G^{d}_{e\tau}/G_{F}.
\EN
The laboratory bounds on the new coupling constants are rather weak.
E.g. from $\nu_e$ scattering we have limits on $\varepsilon=
\sqrt{\varepsilon_e^2+\varepsilon_{e\tau}^2}$ parameter:
$-2.73 <\varepsilon <0.81$ (without fixing axial-vector coupling $\xi$)
$-1.10<\varepsilon<0.64$ ($V+A$ coupling, $\xi=1$) and
$-0.14<\varepsilon<0.15$ ($V-A$ coupling, $\xi=-1$) \cite{BPW3},
whereas there is no reliable limit on the
$\varepsilon_{\tau}$.

Let us assume also that neutrino mass terms are negligible. Then the
Hamiltonian of neutrino evolution in matter takes the form:
\EQ
\hat{H}_{\nu}=\frac{\sqrt{2}G_{F}\rho}{m_{N}}\left(
\begin{array}{cc}
v_e & v_{e\tau}\\
v_{e\tau} & v_{\tau}
\end{array}\right)
\EN
where, recalling that $Y_d=Y_e+2Y_n=1+Y_n$, we have:
\EQ
v_e=1-\frac{3}{2}Y_{n} +\varepsilon_{e}(1+Y_{n}), ~~~~
v_{e\tau}=\varepsilon_{e\tau}(1+Y_{n}), ~~~~
v_{\tau}=-Y_{n}/2 +\varepsilon_{\tau}(1+Y_{n})
\EN
(For the matter antineutrino states $\hat{H}_{\tilde{\nu}}$ is just
distinguished by opposite sign.)
Obviously, due to the new flavour changing interactions an effective mixing
appears between neutrino matter eigenstates
$\nu_{1m} = c_{m}\nu_{e}+s_{m}\nu_{\tau}$ and
$\nu_{2m} = -s_{m}\nu_{e}+c_{m}\nu_{\tau}$:
\EQ
 c_{m} = \mbox{cos}\theta_{m},~~~~ s_{m}=\mbox{sin}\theta_{m},~~~~~
\mbox{tg}2\theta_{m}=\frac{2\varepsilon_{e\tau}}
{(\varepsilon_{\tau}-\varepsilon_{e})-\frac{1-Y_{n}}{1+Y_{n}}}
\EN
The mixing angle between antineutrino matter eigenstates
$\tilde{\nu}_{1m}$
 and $\tilde{\nu}_{2m}$ is the same. For the Hamiltonian eigenvalues we find
\EQ
V_{1,2} = \sqrt{2}G_{F}\frac{\rho}{m_{N}} \,v_{1,2}
\EN
\begin{displaymath}
v_{1,2}= \frac{1}{2}
\left(v_e+v_{\tau} \mp \frac{|v_{\tau}-v_e|}{v_{\tau}-v_e}
\sqrt{(v_{\tau}-v_e)^{2} +4v_{e\tau}^{2}}\right).
\end{displaymath}
and for antineutrinos $V_{\tilde{1},\tilde{2}}= -V_{1,2}$.
Then the transition matrix between $\nu-\tilde{\nu}$ matter eigenstates becomes
\EQ
(\bar{\nu}_{1m}~\bar{\nu}_{2m})\left(
\begin{array}{cc}
h_{11}
 & h_{12}\\
h_{12} &
h_{22}
\end{array}\right)\frac{i}{2}\chi
\left(
\begin{array}{c}
\tilde{\nu}_{1m}\\ \tilde{\nu}_{2m}
\end{array}\right)
\EN
Where
\EQ
h_{11} = s^{2}_{m}h_{\tau\tau},~~~~ h_{12} = -c_{m}s_{m}
h_{\tau\tau},~~~~ h_{22} = c^{2}_{m}h_{\tau\tau}
\EN
(contributions from $h_{ee}, h_{e\tau}$
are neglected because of the strong limits of eqs.(23)).
The key point now is that if the matter mixing angle
$\theta_{m}$ is not very small\footnote{In fact, it is expected to be
reasonably large,
if the flavour-changing interaction has the strength comparable to Fermi
constant, i.e. $\varepsilon_{e\tau}\sim 1$.} the large coupling constant
$h_{\tau\tau}$ can propagate to every entry of the majoron transition
matrix (31) between $\nu$ and $\tilde{\nu}$ matter eigenstates.
Thus, the transitions $\nu_{i}\rightarrow\tilde{\nu}_{j}+\chi$ or
$\tilde{\nu}_{i}\rightarrow\nu_{j}+\chi$ are possible
if are allowed by positive phase space.
(We remind that in matter only helicity-flipping decays are relevant
for very light neutrinos.)
Since in the solar medium the $\nu_e=c_m\nu_{1m}-s_m\nu_{2m}$
state is produced, we are interested only in decays of
neutrino mass eigenstates $\nu_i\rightarrow\tilde{\nu}_j+\chi$
($i,j=1,2$), for which the widthes are:
\EQ
\Gamma_{i\tilde{j}}= \frac{h^{2}_{ij}}{16\pi} \, (V_i+V_j) \,
\Theta(V_i+V_j)
\EN
where the $\Theta$-function remarks that the decay occurs only when the
relevant phase space is positive.
In order to discuss the implication of this scenario for the solar neutrino
problem, we need the expression of the decay probability.
As was mentioned above, the energy independence implies that the decay
probability in a medium with varying density is essentially determined
by the effective matter width passed by neutrinos. This is not absolutely
exact in our case, since the effective coupling constants $h_{ij}$
are also variable during neutrino propagation in the sun.
Apart from this, the new non-standar interactions (24) will also contribute
to matter effective widthes. Moreover, now neutrino mixing angle
also varies during propagation. Even in absence of decay, this implies
the possibility of matter-induced oscillations (up to resonant
conversion \cite{Roulet,BPW3}) which should be taken into account.
As a result of
both matter oscillation and decay effects, the initial $\nu_e$ flux is
converted into $\nu_e$ and $\nu_{\tau}$ fluxes having the initial energy
spectrum (due to the oscillation effect) and $\tilde{\nu}_{e}$ and
$\tilde{\nu}_{\tau}$ fluxes strongly degraded in the energy spectrum (due
to the MID).\footnote{In order
not to intefere with resonant neutrino conversion and, thereby,
not to provide over-suppression of solar neutrino flux, we should
exclude the interval of non-standard interactions
which implies the existence of resonance at sufficiently large densities.
According to ref. \cite{BPW3}, the interval
$\varepsilon_{\tau}-\varepsilon_{e}= 0.5\div 0.75$ is relevant for MSW
conversion (namely, at the lower limit fully adiabatic conversion happens,
at the upper one moderate non-adiabatic regime occurs).}

Using the fact that the decay probabilities do not depend on the neutrino
energy, we give now the exact analytical expressions for the expected
fluxes of $\nu_e, \nu_{\tau}, \tilde{\nu}_e$ and $\tilde{\nu}_{\tau}$ at
the earth:
\begin{displaymath}
\Phi_{\nu_{e}}(E) = \Phi_{SSM}(E)\int_{0}^{R}\int_{0}^{1}
dr d(\mbox{cos}\theta') c^{2}_{m}\left(r\right)
Q\left(r\right)e^{-P_{1}(t,R)}
\end{displaymath}
\begin{displaymath}
\Phi_{\nu_{\tau}}(E) = \Phi_{SSM}(E)\int_{0}^{R}\int_{0}^{1}
dr d(\mbox{cos}\theta')s^{2}_{m}\left(r\right)
Q\left(r\right)e^{-P_{2}(t,R)}
\end{displaymath}
\begin{displaymath}
\Phi_{\tilde{\nu}_{e}} (E) = \int_{0}^{R}\int_{0}^{1}
drd(\mbox{cos}\theta') Q(r)\left[c^{2}_{m}\left(r\right)
B_{1\tilde{1}}(t)+
s^{2}_{m}\left(r\right)
B_{2\tilde{1}}(t)
\right]\tilde{\Phi}(E)
\end{displaymath}
\EQ
\Phi_{\tilde{\nu}_{\tau}}(E) =\int_{0}^{R}\int_{0}^{1}
drd(\mbox{cos}\theta') Q(r)\left[c^{2}_{m}\left(r\right)
B_{1\tilde{2}}(t)+
s^{2}_{m}(r)B_{2\tilde{2}}
(t)
\right]\tilde{\Phi}(E)
\EN
where
\begin{displaymath}
\tilde{\Phi}(E)=2 \,\int_{E}^{E_{end}}dE'
\frac{E'- E}{E'^{2}}\Phi_{SSM}(E'),~~~~t=r\mbox{sin}\theta',
\end{displaymath}
\begin{displaymath}
P_{i\tilde{j}}(t,l) =
\int_{0}^{l} dl'\Gamma_{i\tilde{j}}(z),~~~
B_{i\tilde{j}}(t)=\int_{0}^{R} dl' \, \frac{\Gamma_{i\tilde{j}}^{2}(z)}
{\Gamma_{i}(z)} \, \mbox{e}^{-P_{i}(t,l')},~~~z=\sqrt{t^{2}+l'^{2}},
\end{displaymath}
\begin{displaymath}
P_{i}= \sum_{\tilde{j}}P_{i\tilde{j}},~~~~
\Gamma_{i}=\sum_{\tilde{j}}\Gamma_{i\tilde{j}}
{}~~~~(i,j=1,2)
\end{displaymath}
and $\Phi_{SSM}(E)$ is the differential flux of $\nu_{e}$'s as expected from
the SSM \cite{B}.

  Therefore, provided that $h_{\tau\tau}\simeq 10^{-1}$ and mixing in matter
is large ($\varepsilon_{e\tau}\sim 1$), the MID can be relevant for the SNP.
Moreover, it can naturally explain the origin of the hierarchy
(4) between the signals of different experiments.
The effective matter widthes given by eqs. (22)
provide some numerical insight of why the lower energy neutrinos
(solar $pp$ neutrinos) are less depleted due to the MID
compared to the higher energy
ones ($^{7}Be$ and $^{8}B$ neutrinos), which in turn explains why we observe
$Z_{Ga}>Z_{K,Cl}$.
On the other hand, some difference between $Z_{K}$ and
$Z_{Cl}$ can be achieved due to neutral current contributions from
$\nu_{\tau}$ and $\bar{\nu}_{e,\tau}$ to Kamiokande events. It is
clear, however, that the difference between Kamiokande and Homestake
signals cannot be very large:
\EQ
Z_K-Z_{Cl}<0.15(1-Z_{Cl})
\EN
Where the upper bound actually corresponds to the limit when the
contribution of the MID mechanism is not relevant and the SNP solution
is due to massless neutrino oscillation in matter due to new
flavour-changing interactions \cite{Roulet,BPW3}. However, taking
into account the possible effects of $\tau$-neutrino mass, the
energy dependence of decay probabilities can be achieved, which is
necessary to split more Kamiokande and Homestake signals.
In general case the magnitudes of the fluxes (34) depend on many parameters,
as are $\varepsilon_e$, $\varepsilon_{\tau}$, $\varepsilon_{e\tau}$,
$h_{\tau\tau}$ and, possibly, $m_{\nu_{\tau}}$,
so that the detailed quantitative study with
selection of the parameter range relevant for the SNP solution deserves
special numerical computations and will be presented elsewhere.

One of the remarkable effects of the neutrino
matter induced decay is the appearence of the solar antineutrino flux.
According to eqs. (34) substantial portion (up to 25 per cents) of solar
neutrinos can be transformed into $\tilde{\nu}_{e}$'s. Due to the
substantial energy degradation there is no contradiction with limits on
$\tilde{\nu}_{e}$ flux from Kamiokande \cite{Barbieri} and LSD
\cite{Saavedra} data. However, this $\tilde{\nu}_{e}$-signal
can be detected in free proton rich detectors as are BOREXINO
or BOREX through the inverse
$\beta$ decay $\tilde{\nu}_{e}+p\rightarrow n+e^{+}$.
E.g., for 100 t of fiducial volume of BOREXINO with positron energy threshold
$E_+=3.7$ MeV one can expect up to 35 events per year, whereas
background due to nearby nuclear power reactors is about 3-4 events
per year \cite{Borex}. (For the comparison, in \cite{BMR} we have shown
that the upper limit on the possible  $\tilde{\nu}_{e}$-signal in the
absence of new flavour-changing interactions can hardly exceed the
level of background). On the other hand, the strong energy degradation
of $\tilde{\nu}_e$'s can discriminate the neutrino decay from the
alternative $\tilde{\nu}_e$-signal provided by hybrid models of
neutrino oscillation and spin-flavour precession \cite{Jenia}, in which
case the $\tilde{\nu}_e$ spectrum should not be significantly altered
as compared to the initial solar $\nu_e$ spectrum.

For testing the MID solution to the SNP is also important to measure
neutral current signal from any $\nu_x$ states in which
the missing solar neutrinos could be transformed. This can be done by new
detectors like SNO \cite {SNO} and BOREX \cite{BOREX}.
Obviously, for any mechanism of solar neutrino conversion
$\nu_e\rightarrow\nu_x$, not changing the initial neutrino energy
(like oscillation or magnetic moment transition into active neutrino or
antineutrino states $\nu_x$), one has to expect the following sum rule:
\EQ
\Phi_{\nu_e}(E)+\sum_{x}\Phi_{\nu_x}(E)=\Phi_{SSM}(E)
\EN
for any energy $E$. As for the neutrino decay scenarios, the energy
degradation of secondary neutrinos implies, that l.h.s. of the eq. (36)
should be less than $\Phi_{SSM}(E)$ for the high energy part of neutrino
spectrum and larger for the low energy fraction (the latter, however,
is rather difficult to observe experimentally).
Taking into account that the SNP
solution through the neutrino decay in vacuum is strongly disfavoured
by combined data of all experiments under operation, the observation
of such a "particle number non-conservation" could strongly point out
the MID solution. Clearly, in the case of oscillation or spin-flavour
precession into sterile states one also expects that
$\Phi_{\nu_e}(E)+\sum_{x}\Phi_{\nu_x}(E)<\Phi_{SSM}(E)$, but now this
inequality will be respected for every part of the spectrum.

Thus, as we see, MID can provide well testable solution to the SNP.

\vspace{0.3cm}

{\bf Acknowledgements.}
We are grateful to G.Fiorentini and M.Vysotsky for valuable discussions
and for reading the manuscript.

\end{document}